\documentstyle[11pt]{article}
 
\def\pa{\partial}
 
\def\g{\gamma} \def\G{\Gamma}
\def\a{\alpha} 
\def\b{\beta} 
\def\d{\delta}

\def\l{\lambda} 
\def\m{\mu} 
\def\n{\nu}

\def\s{\sigma}

\def\be{\begin{equation}}
\def\ee{\end{equation}}
\def\mn{{\mu\nu}}
     
\renewcommand{\baselinestretch}{1.0}

\begin{document}

\begin{flushright}
BRX TH-430
\end{flushright}

\vspace*{.2in}

\begin{center}
{\LARGE\bf Born--Infeld--Einstein Actions?}

\vspace{.2in}

S. Deser\footnote{\tt deser@binah.cc.brandeis.edu;
$^2$G.W.Gibbons@damtp.cam.ac.uk}\\
Department of Physics, Brandeis University\\
Waltham,  Massachusetts 02254, USA

\vspace{.1in}

and

\vspace{.1in}

G.W. Gibbons$^2$\\
D.A.M.T.P., Cambridge University \\ 
Silver Street, Cambridge CB3 9EW, UK
\end{center}

\renewcommand{\baselinestretch}{2.0} 
\small
\normalsize                    
\begin{abstract}

We present some obvious physical requirements on
gravitational avatars of non-linear electrodynamics
and illustrate them with explicit determinantal
Born--Infeld--Einstein models. A related procedure,
using compensating Weyl scalars, permits us to formulate
conformally invariant versions of these systems as 
well.
\end{abstract}

Born--Infeld (BI) electrodynamics \cite{001} has earned
its longevity through its elegant, compact, determinantal
form,
\be                                                 
I_{BI} = -1/2 \, \lambda^2 \, \int d^4x \,
\left\{ -| g_{\mu\nu} + \lambda F_{\mu\nu} |
\right\}^{\frac{1}{2}} \; .
\label{eq:1}
\ee
It reduces to Maxwell theory for small amplitudes and
shares with it two special properties, duality invariance
\cite{002} and causal, physical propagation \cite{003}.
Its quartic terms reproduce the effective action of
one-loop SUSY QED.  Not surprisingly, it regularly
surfaces in more general contexts, most recently in 
various aspects of strings, branes and
M.  A further asset
of (\ref{eq:1}) is the absence of ghost photon
modes that are associated with models involving
explicit derivatives on quadratic terms.  This
means that its famous taming of the Coulomb self-energy
is not obtained at the price of ghost compensation,
but really stems from its nonpolynomial 
nature\footnote{One must distinguish, however,
between ``non-singular", BIonic, finite energy 
and ``solitonic": there is no
singularity in the sense that the ``true" Coulomb
field is bounded, but it is still generated by a
point charge in a normal Poisson equation.}
and concomitant dimensional constant $\lambda$.

Is there a gravitational analog of BI? This note is 
part of the problem rather than
of the solution: we will merely present and illustrate,
but without giving
any compelling examplar, some criteria that it
would have to satisfy.

We begin with the purely gravitational sector.
Determinant forms of gravity have a long history 
of their own,
although in a spirit different from that of
BI. The first such action was given by
Eddington \cite{004} in a remarkably ``modern"
spirit; the metric enters as an integration
constant in solving the equations of an ostensibly
purely affine action,
\be
I_{EDD} = \int d^4x |R_{(\mn)} (\G )|
^{\frac{1}{2}}
\label{eq:2}
\ee
where the independent field is a symmetric affinity
$\G^\a_{\n\m} = \G^\a_{\mn}$, and $R_{(\mn )}$
the symmetric part of its 
(generically nonsymmetric) Ricci tensor $R_{\mn}$.
Its (purely ``Palatini") variation implies that
the covariant gradient of the normalized
minor of $R_{(\m\n)}$ vanishes.  Consequently,
$R_{(\mn )}$ is a ``metric" for the affinity,
$R_{(\mn )} = \l g_{\mn}, \; D_\a (\G ) g_{\mn}=0$.
This model has given rise to large literature of its
own, including such extensions as using the full 
Ricci tensor and nonsymmetric $\G$ to represent 
electromagnetism.  Although we hope to
return to these aspects, it is not the road 
we take here. Ours is closer to the spirit of
\cite{001}, working
with a metric manifold from the start, with 
the generic geometrical action
\be
I_G = \int d^4x  \{ -| a g_{\mn} + bR_{\mn} +
cX_{\mn} |\}^{\frac{1}{2}} \; .
\label{eq:3}
\ee
We have separated the linear Ricci term from terms
$X_{\mn} (R)$ quadratic or higher in curvature.
The major necesary condition here is that (\ref{eq:3})
describe gravitons but no ghosts.  This simply means
that the curvature expansion of (\ref{eq:3}) should
begin with the Einstein $R$ (of proper sign of
course) but not contain any quadratic terms, since
the latter are always responsible for ghost modes in
an expansion about Minkowski (or de Sitter) 
backgrounds.  [The cosmological term implicit in 
(\ref{eq:3}) (and indeed in (\ref{eq:2}))
can always be removed by subtraction
or by suitable parameter limits.]  To understand
the effect of this constraint, we recall that
in D=4, the Gauss--Bonnet combination
\be
E_4 \equiv \sqrt{-g} \, 
[R^2_{\mn \a\b} - 4R^2_\mn + R^2 ]
\label{eq:4}
\ee
is a total divergence, so that effectively the
generic quadratic Lagrangian is 
$\a R^2_\mn + \b R^2$; since these terms
(in any combination)
always generate ghosts or tachyons, they must be 
absent.\footnote{Also in the BI spirit, we exclude
explicit derivatives in $X_\mn$, although 
ghost-freedom alone does not exclude them in cubic
or higher terms.}  In fact, the quadratic part of
$E_4$ in $h_\mn \equiv g_\mn - \eta_\mn$ is a
total divergence in any $D$, so the above remarks
remain valid there.  In (\ref{eq:3}) we could also
have used the more general combination
$\tilde{R}_\mn = R_\mn - ag_\mn R, \; a \neq 1/4$,
including even $\tilde{R}_\mn \equiv g_\mn R$;
this trivializes the BI procedure, 
resembling a choice
$\sim | g_\mn (1-\l^2 F^2)|^{\frac{1}{2}}$ there.  The
expansion of a determinant, 
\be
|1 +A|^{\frac{1}{2}} = 1 + \frac{1}{2} \:
tr \: A + \frac{1}{8} (tr\, A)^2 - \frac{1}{4} \,
tr (A^2) + {\cal O} (A^3)
\label{eq:5}
\ee
tells us that we must cancel the terms 
$\sim \frac{b^2}{8} (\frac{1}{2} \, R^2 - R^2_\mn )$
due to the quadratic expansion in $R_\mn$ by using the
leading parts of $\frac{1}{2} \, X^\a_\a$.  This leaves 
a wide latitude in the choice of $X_\mn$: firstly, we can 
use any $X_\mn$ whose trace 
$\sim - \frac{b^2}{8} (\frac{1}{2} \, R^2 - R^2_\mn )
+ f E_4$  for arbitrary constant $f$.  We can then
obviously arrange for this $X^\a_\a$ value by having
$X_\mn$ be a pure trace term, 
$X_\mn = \frac{1}{4} \, g_\mn X^\a_\a$, or with more
exotic choices such as $X_\mn \sim (R_\mn R^\a_\n -
\frac{1}{2} \, RR^\a_\a )$ or any (suitably normalized)
linear combinations of these.  Their differences will
only show up in cubic, ${\cal O}(XR)$, and higher 
contributions. There is no immediate criterion, 
obtainable from ghost-freedom, to further constrain
$X$, although one may imagine adding higher and higher
powers of $R_{\mn}$ in $X$ to cancel particular 
unwanted contributions from expanding the $R_\mn$ and
mixed contributions order by order.

In this connection, let us note that a ``fudge tensor"
$X_\mn$ actually permits one to write almost any 
action in BI form, so that there must be some 
{\it a priori} criterion for it as well.  For example,
any electromagnetic Lagrangian $L_0 (\a ,\b )$, where
$\a \equiv \frac{1}{2} F^2_\mn$,
$\b \equiv \frac{1}{4} F^*F$ are the two invariants,
can be so expressed: Simpy write $f\equiv L_{BI}/L_0$ and
factorize $f^{-\frac{1}{2}}$ into each element,
$(f^{-\frac{1}{2}}g_\mn + \l f^{-\frac{1}{2}}
F_\mn)$ of the new determinant that now represents 
$L_0$, then expand $f$ about unity and call the rest
$X_\mn$.

There are further possible criteria: 
one may require that there
be no singular ``Coulomb" -- here
Schwarzschild (or Schwarzschild--de Sitter ones if the
cosmological constant is kept) -- solutions.  
This in turn has the
necessary consequence that Ricci-flat solutions are to be
excluded, meaning that at some order in the field equations
there must appear terms depending only on the Weyl tensor, as
can be accomplished simply by endowing $X_\mn$
with Weyl 
tensor dependence.\footnote{This condition is not sufficient,
as one could imagine combinations of such terms in
the field equations that vanish for the simple
Schwarzschild (or similar) form but not for generic
Ricci-flat spaces.  Of course certain Einstein spaces
will remain solutions of any action, namely the
(unbounded) $pp$-waves, all of whose scalar invariants
vanish \cite{005}; electromagnetic plane waves are
likewise solutions of any nonlinear model, since
their $\a$ and $\b$ vanish.}  Presumably the space of such 
theories will be further
constrained by the requirement that {\it their}
``Coulomb" solutions will be milder than the black
holes they replace!  Perhaps the strongest ``physical" 
constraint on theories of this type, however, 
would come from demanding that (like BI) they be
supersymmetrizable.  We do not know even know
if this is at all
possible, since the SUSY would have to be a local
one, a very stringent (and dimension-dependent)
requirement.  The positive-energy issue might 
constitute one major barrier. Our other conditions are 
only mildly dimension-dependent: Although for D$>$4,
the Gauss--Bonnet identity is replaced by higher
curvature ones (in even $D$) that are irrelevant to
the ghost problem, we have
seen that the linearized D=4 identity
is preserved in any $D$, so
the D=4 discussion effectively stands. For D=3, $E_4 
\equiv 0$, Riemann and Ricci tensors
coincide so the exercise would reduce to some
variant of Einstein theory. 
For D=2 only the scalar Euler density $R$ remains
and no ``genuine" BI
structure is possible, although one can still write
\cite{006} a
form  $\sim |g_\mn \, f(R)|^{\frac{1}{2}}$.
The original BI action (\ref{eq:1}) is of course 
insensitive to $D$.

Let us now turn to possible ``BI-E" actions involving
both photons and gravitons.  Reinstating the
linear $\l \, F_\mn$
term inside our determinant in (\ref{eq:3}) will now give
rise to
nonminimal cross-terms at least bilinear in $F$
(because it is antisymmetric) times powers of curvatures.
While such terms do not affect the excitation content,
they do alter the propagation properties of both types
of particles \cite{007}.  Indeed their ``light cones" become 
governed by effective metrics of the schematic form
$(g_\mn + {\cal O}_\mn )$ where ${\cal O}_\mn$ represents
the nonminimal contribution, with attendant propagation
complications that may in priciple be used to narrow
the ambiguity in $X$.  Clearly, one would also want
a suitably tamed Riessner--Nordstrom solution here,
along with a bounded Coulomb field. Scalar fields 
(or multiplets) can be incorporated
in a ghost-free way by adding terms of the form
$\sim (\pa_\m \phi \pa_\n \phi + m^2 g_\mn \phi^2)$
under the determinant; their trace will reproduce
the usual scalar action, but now there will be 
non-minimal cross-terms as well.

\section{Weyl-invariance } 

In four spacetime dimensions, the Maxwell action is 
invariant under Weyl rescalings of the metric:
\be
g_{\mu \nu} \rightarrow \Omega ^2 (x) g_{\mu \nu}
\label{eq:6}
\ee
without transforming the $A_\m$.
The Born-Infeld action is of course not Weyl-invariant.
However, it may  readily be made so by introducing a 
scalar compensating field
$\phi$ of Weyl weight $-1$,
\be
\phi \rightarrow   \Omega^{-1} \phi \;.
\label{eq:8}
\ee
The modified action is
\be                                                 
I_{BIW} = -1/2 \, \lambda^2 \, \int d^4x \,
\left\{ -| \phi ^2 g_{\mu\nu} + \phi ^{-2} 
{\cal D}_\mu \phi {\cal D}_\nu
 \phi  + \lambda F_{\mu\nu} |
\right\}^{\frac{1}{2}} \; ,
\label{eq:9}
\ee
where the (real!) Weyl-covariant derivative is
${\cal D}_\mu = \partial _\mu + A _\mu$ 
and the vector potential now
undergoes an ${ R ^\star}$
gauge transformation:
\be
A_\mu \rightarrow  A_\mu + 
\partial_\m \; ln \; \Omega \;. 
\label{eq:10}
\ee 
An appropriate choice of sign of the scalar term
ensures its ghost-freedom.

This model can also accomodate Weyl invariant gravity,
though of course that always involves ghosts.  In
particular one could add 
in (\ref{eq:9}) a (fourth derivative) 
combination of the form 
\be
a\phi^2 g_{\m\n} C^\a_{~\b\g\d}\; C^\b_{~\a\l\s}\;
g^{\g\l}g^{\d\s} +
b\phi^2 C_{\m~\b\d}^{~\a} \;
        C_{\n~\a\s}^{~\b} \; g^{\d\s} 
\label{eq:11}
\ee
(for D=4 the two terms in (\ref{eq:11})
are  proportional).  An alternate route is to 
``improve" the Einstein
term using the compensator: the relevant
(D=4) action is  proportional to the
famous combination
$\phi^{-2} \sqrt{-g} (\frac{1}{6}\, R + 
\phi^{-2} (\partial_\m \phi \partial_\n \phi )
g^{\mn} )$ where either the Einstein or the
gravity kinetic term is now necessarily of the wrong
sign \cite{008}.  This model can obviously
be adapted to BI form
using the previously discussed extensions of 
$R_\mn$ and $X_\mn$.

In conclusion, it should be obvious from the 
rather loose conditions
we have stated that any real progress on adding
``E" to BI will require either better hints from
string expansions or from supersymmetry requirements.
The elegant pure BI insight has as yet found no 
counterpart here. 

After this work was completed, an interesting
non-determinantal two-metric
reformulation of BI has been suggested \cite{009}.
In the process, evenness of BI in $F_{\mn}$ is used
to rewrite the determinant (\ref{eq:1}) in terms
of the tensor $(g_{\m\n} + \l^2 F_{\m\a}F^\a_\n )$.
While it is perhaps formally more natural to
include $R_\mn + X_\mn$ into this symmetric array,
our considerations remain unaltered.

The research of S.D. was supported by 
the National Science Foundation, under
grant \#PHY-9315811. Our work was begun 
at the VI Conference on
the Quantum Mechanics of Fundamental 
Systems, during a session held at the Presidente 
Frei Antarctic base;  we thank the organizers 
for this unique opportunity.

\end{document}